\documentclass[a4paper]{jpconf}
\usepackage{graphicx}

\begin{document}
\title{Low Background Micromegas in CAST}

\author{J~G~Garza$^1$,  S~Aune$^2$, F~Aznar$^{1,}$\footnote[7]{Present address: Centro Universitario de la Defensa, Universidad de Zaragoza, Ctra.~de Huesca s/n, 50090 Zaragoza, Spain}, D~Calvet$^2$, J~F~Castel$^1$, \linebreak F E Christensen$^{4}$, T~Dafni$^1$, M~Davenport$^5$,  T~Decker$^{3}$,  E~Ferrer-Ribas$^2$,  J~Gal\'an$^2$, J~A~Garc\'ia$^1$, I~Giomataris$^2$, R~M~Hill$^{3}$, F~J~Iguaz$^1$, I~G~Irastorza$^1$, A~C~Jakobsen$^{4}$, D~Jourde$^2$, H~Mirallas$^1$, I~Ortega$^{1,}$\footnote[8]{Present address: CERN, European Organization for Particle Physics and Nuclear Research}, T~Papaevangelou$^2$, M~J~Pivovaroff$^{3}$, J~Ruz$^{3}$, A~Tom\'as$^{1,}$\footnote[9]{Present address: High Energy Physics group, Brackett Laboratory, Imperial College, London, UK}, T~Vafeiadis$^5$, J~K~Vogel$^{3}$}

\address{$^1$ Grupo de F\'isica Nuclear y Astropart\'iculas, University of Zaragoza, Zaragoza, Spain}
\address{$^2$ Irfu, CEA, Centre de Saclay, Gif sur Yvette, France}
\address{$^3$ Lawrence Livermore National Laboratory, Livermore, CA, USA}
\address{$^4$ Technical University of Denmark, DTU Space Kgs. Lyngby, Denmark}
\address{$^5$ CERN, European Organization for Particle Physics and Nuclear Research, Geneva, Switzerland}

%\address{$^3$ Institute of Nuclear Physics, NCSR Demokritos, Athens, Greece} 
%\address{$^4$ University of Patras, Patras, Greece}

\ead{jgraciag@unizar.es}

\begin{abstract}
Solar axions could be converted into x-rays inside the strong magnetic field of an axion helioscope, triggering the detection of this elusive particle. Low background x-ray detectors are an essential component for the sensitivity of these searches. We report on the latest developments of the Micromegas detectors for the CERN Axion Solar Telescope (CAST), including technological pathfinder activities for the future International Axion Observatory (IAXO). The use of low background techniques and the application of discrimination algorithms based on the high granularity of the readout have led to background levels below 10$^{-6}$~counts/keV/cm$^2$/s, more than a factor 100 lower than the first generation of Micromegas detectors. The best levels achieved at the Canfranc Underground Laboratory (LSC) are as low as 10$^{-7}$ counts/keV/cm$^2$/s, showing good prospects for the application of this technology in IAXO. The current background model, based on underground and surface measurements, is presented, as well as the strategies to further reduce the background level. Finally, we will describe the R\&D paths to achieve sub-keV energy thresholds, which could broaden the physics case of axion helioscopes.
\end{abstract}

%%%%%%%%%%%%%%%%%%%%%%%%%%%%%%%%%%%%%%%%%%%%%%%%
%%%%%%%%%%%%%%%%%%%%%%%%%%%%%%%%%%%%%%%%%%%%%%%%
% Micromegas in CAST
%%%%%%%%%%%%%%%%%%%%%%%%%%%%%%%%%%%%%%%%%%%%%%%%
%%%%%%%%%%%%%%%%%%%%%%%%%%%%%%%%%%%%%%%%%%%%%%%%
\section{Micromegas for axion searches}
\label{sec:mm}

Axions were proposed to solve the strong-CP problem more than 35 years ago by R.~D.~Peccei and H.~R.~Quinn\cite{cp}. Now, they are still the most compelling solution, and  one of the few viable candidates to compose the dark matter. Axion helioscopes \cite{sikivie, bibber} aim to detect solar axions by their conversion into x-rays (1-10 keV) in the presence of strong magnetic fields. The CERN Axion Solar Telescope (CAST)\cite{cast,limit1, limit2} is the most powerful implementation of this technique so far, setting the most stringent limits on the axion-photon coupling constant $g_{a\gamma}$ for a wide range of axion masses \cite{limit1, limit2}. The future International Axion Observatory (IAXO) aims to increase the sensitivity to the axion-photon coupling constant by 1-2 orders of magnitude  \cite{iaxo, cdr}.

The sensitivity of an axion helioscope can be considered in terms of four main ingredients (magnet, optic, x-ray detector, exposure) of such an experiment as $g_{a\gamma} \propto b^{1/2}\epsilon^{-1} \times s^{1/2}\epsilon_{o}^{-1} \times (BL)^{-2}A^{-1} \times t^{-1/2}$ \cite{ngah}. Here, $t$ is the exposure time, $L$ and $A$ are the length and cross-sectional area of the magnet, $B$ the intensity of the magnetic field, $s$ and $\epsilon_{o}$ are the focusing spot area and efficiency of the x-ray optics, and $b$ and $\epsilon$ are respectively the background level and signal efficiency of the x-ray detectors.

\begin{figure}
\begin{center}
\includegraphics[height=0.31\textwidth]{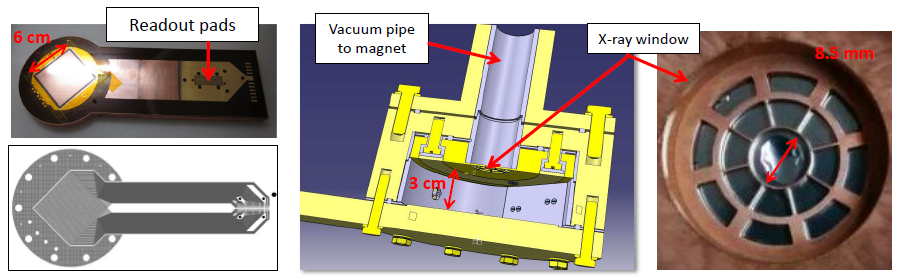}
\end{center}
\caption{The CAST microbulk Micromegas readout parallelizes 6$\times$6 cm in 120$\times$120 strips (left). Detector sketch showing the main components of the detection chamber (center). The x-ray window is a gas-tight, 4 $\mu$m, aluminized mylar foil glued on a spider-web patterned copper strong-back that withstands the differential pressure to the magnet vacuum system (right).  }
\label{imagen1}
\end{figure}

Therefore, high-efficient low-background x-ray detectors are mandatory for an axion helioscope. Microbulk Micromegas (MM) x-ray gaseous detectors \cite{mm} are currently used to equip three of the four CAST magnet bores, and they have been proposed as the baseline detector technology for IAXO (figure \ref{imagen1}). The choice of this technology is based on several properties: a) the readout is highly granular, allowing a strong background suppression based on the different topology of the x-ray signal (point-like) and the gamma and muon background (extended, non-symmetric tracks) \cite{lowmm}; b) all the detector components are carefully chosen to have low radioactivity, including the Micromegas readout, whose radioactivity has been measured in a high-purity germanium detector in the Canfranc Underground Laboratory (LSC)~\cite{radiopure}. c) active and passive shielding techniques can be applied as in rare event searches; d) their manufacture relies on a consolidated technique that guarantees the stability of the readout over long running periods; high and uniform gains are achieved, and good energy resolution (12\%~FWHM at 5.9~keV) can be reached \cite{fwhm, pacoNeon}.

%it is intrinsically radiopure as the readout presents both low material budget and low radioactivity \cite{radiopure}; 

%%%%%% Figure MM %%%%%%%%
%%%%%%%%%%%%%%%%%%%%%%%%

The integration of an x-ray focusing device with a low background MM detector is presented in section \ref{sec:optics}. The latest background levels achieved in CAST-MM detectors \cite{mpgd2013} and in the Canfranc Underground Laboratory make the MM technology an excellent candidate for accomplishing the IAXO requirements, namely $10^{-7}-10^{-8}$ keV$^{-1}$cm$^{-2}$s$^{-1}$ in the energy range of interest (RoI). The current background model and the strategies to further reduce the background level are presented in section \ref{sec:bkg}, while in section \ref{sec:threshold} some research paths for lowering the energy threshold are presented.

%%%%%%%%%%%%%%%%%%%%%%%%%%%%%%%%%%%%%%%%%%%%%%%%
%%%%%%%%%%%%%%%%%%%%%%%%%%%%%%%%%%%%%%%%%%%%%%%%
%% MM and x-ray optics
%%%%%%%%%%%%%%%%%%%%%%%%%%%%%%%%%%%%%%%%%%%%%%%%
%%%%%%%%%%%%%%%%%%%%%%%%%%%%%%%%%%%%%%%%%%%%%%%%
\section{MM and x-ray optics}
\label{sec:optics}

\begin{figure}
\begin{center}
\includegraphics[width=0.9\textwidth]{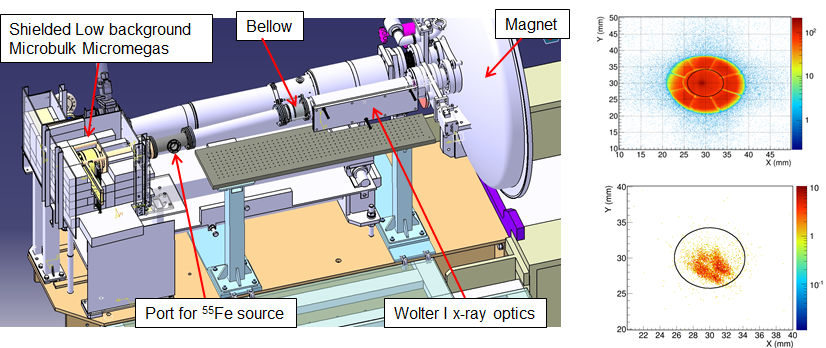}
\end{center}
\caption{Sketch of the CAST detection line composed of a low background MM placed at the focal point of an x-ray telescope (left). Intensity map of a regular $^{55}$Fe calibration (right-up) where the shadow of the strong-back is observed. The same is represented for x-rays coming through the optics from a  source placed 13 m away (right-down).}
\label{imagen2}
\end{figure}

The helioscope technique can be enhanced by the use of an x-ray focusing device, increasing the signal-to-background ratio and thus the sensitivity. A new dedicated x-ray optic was installed at one end of the CAST magnet in 2014, with a low background MM at its focal plane. On top of increasing CAST's sensitivity, it is a demonstration of the techniques proposed in the conceptual design report for IAXO \cite{cdr}. The line is composed of a $\sim$5 cm diameter, 1.3 m focal-length Wolter I x-ray telescope (XRT) and a shielded MM detector made of radiopure components (figure \ref{imagen2}, left). For the first time, the telescope has been specifically designed for axion research using NUSTAR technology~\cite{nustar,xrayoptics}.   

This line was successfully commissioned shortly after its installation at CAST, and it is currently looking for solar axions. Figure \ref{imagen2} (top-right) shows the intensity map registered by the MM detector produced with a regular $^{55}$Fe source placed behind the x-ray optic. The intensity map produced in the MM detector by an x-ray source focused by the telescope is shown in figure \ref{imagen2} (bottom-right): it approximately defines the region where the axion signal is expected. The increase in sensitivity due to the x-ray optic focusing is roughly estimated as a factor $5-8$.  A detailed paper reporting the design, operation and results of the MM+XRT system is in preparation, as well as a more technical paper on the x-ray focusing device.

%(strictly, the source should be placed in the infinite instead of 13 m away from the telescope)

%%% Figure MM-optics
%%%%%%%%%%%%%%%

%%%%%%%%%%%%%%%%%%%%%%%%%%%%%%%%%%%%%%%%%%%%%%%%%%%%%%%
%%%%%%%%%%%%%%%%%%%%%%%%%%%%%%%%%%%%%%%%%%%%%%%%%%%%%%%%
%% background
%%%%%%%%%%%%%%%%%%%%%%%%%%%%%%%%%%%%%%%%%%%%%%%%%%%%%%%
%%%%%%%%%%%%%%%%%%%%%%%%%%%%%%%%%%%%%%%%%%%%%%%%%%%%%%%
\section{CAST-MM background model and strategies for reduction}
\label{sec:bkg}

\begin{figure}
\begin{center}
\centering
\includegraphics[height=0.35\textwidth]{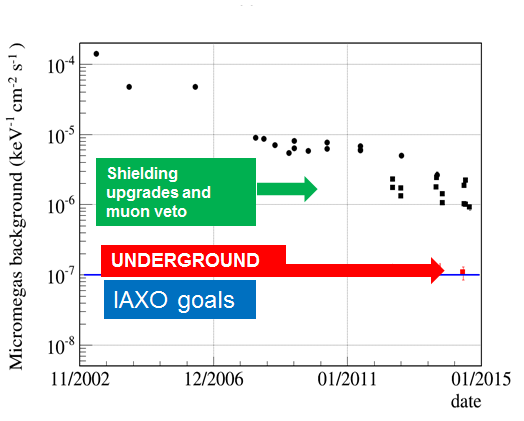} 
\includegraphics[height=0.35\textwidth]{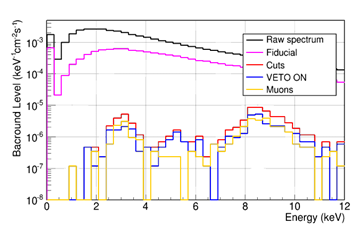} 
\end{center}
\caption{Evolution of the background level in CAST, the limit obtained in the LSC and IAXO goals (left). Background energy spectrum of the CAST line in 2014 (right). }
\label{imagen3}
\end{figure}

The application of background suppression techniques, the reliability of the microbulk technology, the upgrade of the readout electronics and the tuning of the rejection algorithms led to a level reduction of more than two orders of magnitude in CAST-MM detectors over the last ten years (figure \ref{imagen3}) \cite{lowmm, juanan}. The current background level is below 10$^{-6}$ keV$^{-1}$cm$^{-2}$s$^{-1}$ in the CAST RoI \cite{mpgd2013}. The corresponding background energy spectrum is characterized by a fluorescence peak at 8 keV (from copper $K_{\alpha}$ emission), its escape peak at 5 keV and by the argon $K_{\alpha}$ emission line at 3 keV (figure \ref{imagen3}). The operation of a replica of the detector in the LSC sets a level as low as $\sim$10$^{-7}$ keV$^{-1}$cm$^{-2}$s$^{-1}$, almost at the required IAXO levels.

%%%% Figure background
%%%%%%%%%%%%%%

Further background reductions depend on the identification of the source of the remaining events and on the application of the necessary techniques to minimize them. The current understanding of the background sources is summarized in table \ref{table1}. This model is based on in-situ measurements at CAST, tests underground (at the LSC) and at surface, as well as in Geant4 simulations. While the contributions from gamma rays, radon and internal radioactivity have been reduced to negligible levels, the dominant component is caused by cosmic muons and their secondary products, generated after their interaction in the setup components. The origin of the events limiting the LSC performance is unknown. Some hypotheses are the $\beta$-decay of $^{39}$Ar present in the detection medium, neutrons or cosmic activation of the materials.

%%%% Table background
\begin{table}
\centering
\label{table1}
\begin{tabular}{ l|cc|c }
Contribution & \multicolumn{2}{c|}{Level (keV$^{-1}$cm$^{-2}$s$^{-1}$)} & Reduction technique applied\\
 & Before & After & \\
\hline
Gamma flux   	& 7$\times$10$^{-5}$	& Negligible? 		&  Full coverage by 10 cm lead shielding \\
Radon   	& 8$\times$10$^{-7}$	& Negligible 			&  Nitrogen flux inside the shielding \\
Cosmic muons    & 2$\times$10$^{-6}$	& 7$\times$10$^{-7}$ 	&  95\% coverage by an active muon veto\\
Al cathode   	& 5$\times$10$^{-7}$	& Negligible	&  Replacement by an ultrapure copper cathode\\
\hline
LSC limit   	& \multicolumn{2}{c|}{1.1$\times$10$^{-7}$} 	 & $^{39}$Ar?, Neutrons?, cosmic activation? \\
\end{tabular}
\caption{Summary of the different contributions to the background level in the RoI of the CAST-MM detectors and the techniques applied to reduce them (table extracted from \cite{pacotipp}).  A detailed description of the measurements and their associated error is given in \cite{lowmm}.}
\end{table}
%%%%%%%%%%%%%

%For clarity, values are given without errors.

The strategy to reduce background at surface level are based on the installation of thicker and more compact shielding and in a 4$\pi$ enlarged muon veto system. Pushing the lowest underground limit requires a change in the active gas to xenon or neon, or the installation of a neutron shielding. These activities are being developed in the context of the R\&D phase for the IAXO technical design report.

%%%%%%%%%%%%%%%%%%%%%%%%%%%%%%%%%%%%%%%%%%%%%%%%%%%%%%%
%%%%%%%%%%%%%%%%%%%%%%%%%%%%%%%%%%%%%%%%%%%%%%%%%%%%%%%
%% threshold
%%%%%%%%%%%%%%%%%%%%%%%%%%%%%%%%%%%%%%%%%%%%%%%%%%%%%%%
%%%%%%%%%%%%%%%%%%%%%%%%%%%%%%%%%%%%%%%%%%%%%%%%%%%%%%%
\section{Towards lower energy thresholds}
\label{sec:threshold}

The efficiency of the x-ray focusing device shifts the signal to lower energies. Even in the hadronic axion production mechanism, an important fraction of the signal is at sub-keV energies. Moreover, axions could be produced at the Sun via non-hadronic processes at an energy peaking around 1~keV~\cite{redondo}. More exotic searches, like paraphotons or chameleons also peak at very low energies. These facts motivate the use of sub-keV energy threshold detectors.

The R\&D paths for lowering the energy threshold include: the use of more transparent x-ray windows made out of other materials or different geometries than presently used; other gas mixtures with higher gain at the operation point \cite{pacoNeon}; self-triggered electronic readouts to improve the signal-to-noise ratio~\cite{aget}; calibration at lower energies that will provide new analysis tools based on the dependence of the event topology on the x-ray energy (see \cite{pacotipp} for more details).

%%%%%%%%%%%%%%%%%%%%%%%%%%%%%%%%%%%%%%%%%%%%%%%%%%%%%%%
%%%%%%%%%%%%%%%%%%%%%%%%%%%%%%%%%%%%%%%%%%%%%%%%%%%%%%%
%% conclusions
%%%%%%%%%%%%%%%%%%%%%%%%%%%%%%%%%%%%%%%%%%%%%%%%%%%%%%%
%%%%%%%%%%%%%%%%%%%%%%%%%%%%%%%%%%%%%%%%%%%%%%%%%%%%%%%
\section{Conclusions}
\label {sec:conclusions}

The Micromegas detectors in CAST have reduced their background level by more than two orders of magnitude in the last decade, reaching values below 10$^{-6}$ keV$^{-1}$cm$^{-2}$s$^{-1}$. The different measurements performed at CAST and at the LSC have allowed to build a background model with our understanding of the system. The main contributions have been identified and partially or totally reduced. We have presented the open R\&D paths for further reducing the remaining contributions together with new research paths for lowering the energy threshold of the MM detectors. Finally, the best levels achieved in the measurements performed in the LSC motivate the use of this technology in IAXO.

%%%%%%%%%%%%%%%%%%%%%%%%%%%%%%%%%%%%%%%%%%%%%%%%%%%%%%%
%%%%%%%%%%%%%%%%%%%%%%%%%%%%%%%%%%%%%%%%%%%%%%%%%%%%%%%
%\acknowledgments
%%%%%%%%%%%%%%%%%%%%%%%%%%%%%%%%%%%%%%%%%%%%%%%%%%%%%%%
%%%%%%%%%%%%%%%%%%%%%%%%%%%%%%%%%%%%%%%%%%%%%%%%%%%%%%%
\ack{We want to thank our colleagues of CAST for many years of collaborative work in the experiment, and many helpful discussions and encouragement. Part of this work was performed under the auspices of the U.S. Department of Energy by Lawrence Livermore National Laboratory under Contract DE-AC52-07NA27344. We thank R. de Oliveira and his team at CERN for the manufacturing of the microbulk readouts. We also thank the LSC staff for their help in the support of the Micromegas setup at the LSC. The authors would like to acknowledge the use of Servicio General de Apoyo a la Investigaci\'on-SAI, Universidad de Zaragoza. F.~I. acknowledges the support from the \emph{Juan de la Cierva} program of the MINECO. We acknowledge support from the European Commission under the European Research Council T-REX Starting Grant ref. ERC-2009-StG-240054 of the IDEAS program of the 7th EU Framework Program. We also acknowledge support from the Spanish Ministry of Economy and Competitiveness (MINECO) under contracts ref. FPA2008-03456, FPA2011-24058 and EIC-CERN-2011-0006, as well as under the CPAN project ref. CSD2007-00042 from the Consolider-Ingenio 2010 program. Part of these grants are funded by the European Regional Development Fund (ERDF/FEDER). }

%%%%%%%%%%%%%%%%%%%%%%%%%%%%%%%%%%%%%%%%%%%%%%%%%%%%%%%
%%%%%%%%%%%%%%%%%%%%%%%%%%%%%%%%%%%%%%%%%%%%%%%%%%%%%%%
%%%%%%%%%%%%%%%%%%%%%%%%%%%%%%%%%%%%%%%%%%%%%%%%%%%%%%%
%%%%%%%%%%%%%%%%%%%%%%%%%%%%%%%%%%%%%%%%%%%%%%%%%%%%%%%


\section*{References}
\medskip
\begin{thebibliography}{9}
\bibitem{cp} Peccei R D and Quinn H R 1977 {\it Rev. Lett.} {\bf 38} 1440-1443
\bibitem{sikivie} Sikivie P 1983 {\it Phys. Rev. Lett } {\bf 51 } 1415
\bibitem{bibber} van Bibber K {\it et al.} 1989 {\it Phys. Rev.} {\bf D39}
%\bibitem{helioscopes} Dafni T, Iguaz F J {\it et al.} 2014 PoS TIPP2014 130
\bibitem{cast} CAST Collaboration, Zioutas K {\it et al.} 2005 {\it Phys. Rev. Lett. } {\bf 94 } 121301
\bibitem{limit1} CAST Collaboration, Adriamonje S 2007 {\it et al.}{\it  JCAP} {\bf 04 } 010
\bibitem{limit2} CAST Collaboration, Arik E 2014 {\it et al.}{\it Phys. Rev. Lett. } {\bf 112} 091302
\bibitem{iaxo} Irastorza I G {\it et al.} 2013 {\it CERN-SPSC-2013-022, SPSC-I-242 } 
%\bibitem{cdr} Armengaud E {\it et al. }  2014 {\it JINST }{\bf 9 } C12042
\bibitem{cdr} Armengaud E {\it et al. }  2014 {\it JINST }{\bf 9 } T05002
\bibitem{ngah} Irastorza I G {\it et al. }  2011 {\it JCAP }{\bf 1106 } 013
\bibitem{mm} Adriamonje S {\it et al.} 2010 {\it JINST} {\bf 5 } P02001
\bibitem{lowmm} Aunse S {\it et al. } 2014 {\it JINST }{\bf 9} P01001
\bibitem{radiopure} Cebrian S {\it et al.} 2011 {\it Astropart. Phys. }{\bf 34 } 354
\bibitem{fwhm} Cebrian S {\it et al.} 2010 {\it JCAP} {\bf 1010 } 010
\bibitem{pacoNeon} Iguaz F J {\it et al.} 2012 {\it JINST} {\bf 7} P04007
\bibitem{mpgd2013} Garza J G {\it et al.} 2013 {\it JINST } {\bf 8 } C12042
\bibitem{nustar} Harrison F A {\it et al. } 2013 {\it Astrophysical Journal }{\bf 770} 103
\bibitem{xrayoptics} Jakobsen A C {\it et al. } 2013 {\it Proc. SPIE }{\bf 8861 }
\bibitem{juanan} Garcia J G {\it et al.} 2013  {\it J. Phys Conf Ser.} {\bf 460} 012003
\bibitem{redondo} Redondo J 2013  {\it JCAP} {\bf 1312} 008
\bibitem{pacotipp} Iguaz F J {\it et al.} 2014 {\it PoS TIPP2014} 295
\bibitem{aget} Anvar S {\it et al.} 2011 {\it Proc. IEEE Nuclear Science Symp.} pp.~745-749

\end{thebibliography}
\end{document}